\documentclass[aps,twocolumn,superscriptaddress,longbibliography]{revtex4-2}

\usepackage{amsmath}
\usepackage{amssymb}
\usepackage{amsthm}
\usepackage{mathtools}
\usepackage{graphicx}
\usepackage{xcolor}
\usepackage{hyperref}
\usepackage{algorithm}
\usepackage{algpseudocode}
\usepackage{tabularx}
\usepackage{bm}
\usepackage{tikz}
\usepackage[T1]{fontenc}

\newcommand{\be}{\begin{equation}}
\newcommand{\ee}{\end{equation}}

\newcommand{\OO}[1]{\mathcal{O}(#1)}
\newcommand{\paren}[1]{\left( #1 \right)}
\newcommand{\brak}[1]{\left[ #1 \right]}

\newcommand{\abs}[1]{\left| #1 \right|}

\newcommand{\mel}[3]{\left\langle{#1}\middle\vert{#2}\middle\vert{#3}\right\rangle}
\newcommand{\ket}[1]{\left\vert{#1}\right\rangle}

\renewcommand{\Im}{\operatorname{Im}}

\DeclareMathOperator\Tr{Tr}

\newcommand{\intbz}{\int_{\text{BZ}}}
\newcommand{\bz}{\text{BZ}}

\newcommand{\vect}[1]{\bm{#1}}
\newcommand{\vk}{\vect{k}}

\newcommand{\CCQ}{Center for Computational Quantum Physics, Flatiron Institute,
162 5th Avenue, New York, NY 10010, USA}
\newcommand{\CCM}{Center for Computational Mathematics, Flatiron Institute, 162
5th Avenue, New York, NY 10010, USA}

\makeatletter
\newcommand{\multiline}[1]{%
  \begin{tabularx}{\dimexpr\linewidth-\ALG@thistlm}[t]{@{}X@{}}
    #1
  \end{tabularx}
}
\makeatother

\begin{document}

\title{High-order and adaptive optical conductivity calculations using Wannier interpolation}

\author{Lorenzo Van Mu\~noz}
\affiliation{Department of Physics, Massachusetts Institute of Technology, 77
Massachusetts Avenue, Cambridge, MA 02139, USA}
\affiliation{\CCQ}

\author{Jason Kaye}
\affiliation{\CCQ}
\affiliation{\CCM}

\author{Alex Barnett}
\affiliation{\CCM}

\author{Sophie Beck}
\affiliation{\CCQ}

\begin{abstract}
  We present an automatic, high-order accurate, and 
  adaptive Brillouin zone
  integration algorithm for the calculation of the optical conductivity with a
  non-zero but small broadening factor $\eta$, focusing
  on the case in which a Hamiltonian in a downfolded model can be evaluated
  efficiently using Wannier interpolation. 
  The algorithm uses iterated adaptive integration to exploit the localization
  of the transport distribution near energy and energy-difference iso-surfaces,
  yielding polylogarithmic computational complexity with respect to $\eta$.
  To demonstrate the method, we compute the AC optical
  conductivity of a three-band tight-binding model, and are able to resolve the Drude and
  interband peaks with broadening in the sub-meV regime to several digits of accuracy.
  Our algorithm
  automates convergence testing to a user-specified error tolerance, providing
  an important tool in black-box first-principles calculations of electrical
  transport phenomena and other response functions.
\end{abstract}

\maketitle

\section{Introduction} \label{sec:intro}

The potential use of strongly correlated materials in a variety of technological applications, from transparent conductors for photovoltaics and display technologies~\cite{Zhang_et_al:2016} to optoelectronics~\cite{Coulter/Manousakis/Gali:2014, Navarro_et_al:2021}, relies on a detailed understanding of their optical properties.
Given a successful characterization and assigment of spectral features, identifying the ideal material for specific applications becomes a matter of optimizing material properties through band structure control at the meV scale.
To this end, accurate first principles prediction of optical conductivity is an indispensable tool for materials discovery and design.

The optical conductivity sensitively
depends on the structure of Brillouin zone (BZ) integrals involving localized
energy and energy-difference isosurfaces. Its longitudinal part is defined in the Kubo formalism
for the long-wavelength limit ($\bm{q} \to 0$) as~\cite{oudovenko06}
\begin{equation} \label{eq:oc}
    \sigma_{\alpha \alpha'}(\Omega) = N_{sp} \pi e^2 \hbar \int_{-\infty}^{\infty} d\omega \, F_\Omega(\omega) \intbz d^3\vk \, \Gamma_{\alpha \alpha'}(\vk,\omega,\Omega),
\end{equation}
and represents the conductivity in the
direction $\alpha$ due to a field of frequency $\Omega$ polarized in the direction $\alpha'$, for $\alpha, \alpha' \in \{x, y, z\}$.
Here, $N_{sp}$ is the spin
degeneracy, $e$ the electron charge, $\omega$ a frequency variable, and
$\vk$ a momentum wavevector.
The contributing optical transitions for a given temperature $T = 1/(k_B \beta)$ are limited to an energy window defined by the Fermi window function $F_\Omega(\omega) = (f(\omega) - f(\omega+\Omega))/\Omega$, with Fermi-Dirac
distribution $f(\omega) = 1/\paren{1+e^{\beta\hbar\omega}}$.
The $\vk$ integral is taken over the BZ, and the transport distribution is given by
\begin{equation} \label{eq:tdist}
 \Gamma_{\alpha \alpha'}(\vk, \omega, \Omega) = \Tr[\hat{v}_\alpha(\vk) \hat{A}(\vk,\omega)
\hat{v}_{\alpha'}(\vk) \hat{A}(\vk,\omega+\Omega)]\,,
\end{equation}
with velocity operator $\hat{v}_\alpha(\vk) = -\frac{i}{\hbar} [\hat{r}_\alpha, \hat{H}_{\vk}]$
and spectral function $\hat{A}(\vk,\omega) = \frac{i}{2\pi}[\hat{G}(\vk,\omega) -
  \hat{G}^\dagger(\vk,\omega)]$. Here, $\hat{r}_\alpha$ is the position operator,
$\hat{H}_{\vk}$ is the Bloch Hamiltonian,
$\hat{G}(\vk,\omega) = (\hbar\omega - \hat{H}_{\vk}
-\Sigma(\vk,\omega))^{-1}$ is the momentum-dependent single-particle retarded Green's function
(with chemical potential $\mu$ absorbed into $\hat{H}_{\vk}$), and $\Sigma$ is
the self-energy.
We refer to $-\Im \Sigma > 0$ as the scattering rate.

In typical applications, the Hamiltonian is restricted to a small number of bands of interest, a process referred to as downfolding~\cite{oudovenko06}.
This yields a tight-binding model characterized by a small matrix-valued function $H(\vk)$, which is smooth in the basis of
Wannier functions
$\ket{\psi^\text{W}_{j\vk}}$, for $j$ an orbital index~\cite{Marzari/Vanderbilt:1997,Souza/Marzari/Vanderbilt:2001}.
The ``gauge-covariant'' velocity matrix elements are defined as \cite{yates07}
\begin{equation} \label{eq:dipole}
  v_{jj',\alpha}(\vk) = \frac{1}{\hbar}\frac{\partial H_{jj'}(\vk)}{\partial k_{\alpha}} + \frac{i}{\hbar} [H(\vk), \mathcal{A}_{\alpha}(\vk)]_{jj'}.
\end{equation}
In this work we use the Wannier basis, in which $H_{jj'}(\vk) = 
\mel{\psi^{\text{W}}_{j\vk}}{\hat{H}}{\psi^{\text{W}}_{j'\vk}}$ are the 
matrix elements of the Hamiltonian and
$\mathcal{A}_{jj',\alpha}(\vk) =
\mel{\psi^{\text{W}}_{j\vk}}{\hat{r}_{\alpha}}{\psi^{\text{W}}_{j'\vk}}$ are
the matrix elements of the position operator.

The primary difficulty of computing the optical conductivity is
the localization of the transport distribution \eqref{eq:tdist}
along the surfaces on which the Green's functions become nearly singular. 
As a consequence, such calculations often require significantly denser
integration grids than are required to compute total energies and charge
densities~\cite{yates07}.
The near-singularities become sharper as the scattering rate decreases, i.e.,
$\Im\Sigma$ approaches zero, for example in low-temperature calculations.
For sufficiently small scattering rates, adaptive integration grids are
essential to achieving converged results~\cite{assmann16}.
Moreover, since the scattering rate may vary widely across the $(\vk,
\omega)$ domain, algorithms which produce uniform accuracy regardless of the scattering rate are desirable.

This paper extends the integration methods introduced in Ref.~\onlinecite{kaye23_bz},
which focused on the single-particle Green's function, to the optical
conductivity.  There, we presented 
highly optimized uniform integration algorithms for large scattering rates, and
showed that algorithms based on \emph{iterated} adaptive integration outperform
\emph{tree-based} adaptive methods \cite{assmann16} (with polylogarithmic rather than quadratic
scaling in the inverse of the scattering rate), providing access to very small
scattering rates. Here, we describe the more sophisticated structure of the transport distribution, and adapt our methods accordingly. We address questions
such as the optimal order for evaluating integrals, the selection of integration tolerances, and the ``peak-missing''
problem, a tendency of adaptive integration schemes to fail to identify and resolve highly
localized features of the integrand in the extremely small scattering regime.
We obtain robust, automatic algorithms which yield user-controllable high-order
accuracy, for scattering rates below the meV scale. By contrast, in the $0.1-1$~meV range, uniform
grid-based integration methods become intractable, requiring $10^3-10^4$
$\vk$-points per dimension to obtain well-resolved results.

We briefly review existing numerical integration methods for the
optical conductivity. We refer to Ref.~\onlinecite{kaye23_bz} for further references on the broader BZ integration problem.
When Wannier interpolation is available, the
most commonly used BZ integration method is a simple sum over a uniform grid,
often referred to as a Monkhorst--Pack \cite{Pizzi_et_al:2020,wang2006,tsirkin21} or periodic trapezoidal rule \cite{kaye23_bz} grid, which is spectrally accurate due to the periodicity of the integrand. Ref.~\onlinecite{vernes2001} proposed an automatic global refinement algorithm using this quadrature rule. Refs.~\onlinecite{wang2006,tsirkin21}
proposed an adaptive algorithm based on local refinement to uniform sub-grids using the magnitude of the integrand as a refinement heuristic, but this approach is low-order accurate and does not give robust error guarantees.
Ref.~\onlinecite{assmann16} presents a low-order accurate tree-based adaptive quadrature
method based on a tetrahedral mesh.
The optical conductivity has also been calculated using the linear tetrahedron method \cite{Haule_et_al:2005,oudovenko06,ambrosch2006}, but this approach is also low-order accurate and has potential failure modes \cite[App. E]{kaye23_bz}.
To compute the frequency integral, Refs.~\onlinecite{assmann16,Haule_et_al:2005} use
uniform grids, and Ref.~\onlinecite{vernes2001} uses contour integration. 
In \textit{ab-initio} settings in which Wannier interpolation
is not available, the kernel polynomial method is a standard tool
for calculating generic spectral quantities in molecular and disordered systems
\cite{joaoKITE2020,weisse2006}. 
To our
knowledge, this work is the first presentation of an adaptive integration algorithm for the
optical conductivity which is robust, high-order accurate, and well-suited for
non-zero but very low
temperature calculations.

Sec.~\ref{sec:background} contains a brief review of the main ideas presented in
Ref.~\onlinecite{kaye23_bz}, and Sec.~\ref{sec:algorithms} describes our
integration algorithms for the optical conductivity.
In Sec.~\ref{sec:results}, we demonstrate the performance of our methods for a simple three-band tight-binding model on a cubic lattice which includes both intra- and interband hopping amplitudes.
Using a Fermi liquid scaling of the self-energy, we reproduce the expected $T^2$-resistivity related to the Drude intraband
conductivity, and resolve the detailed structure of the interband peak for decreasing temperatures and scattering rates. Our results are accurate to several significant digits at
previously inaccessible temperature and energy scales as low as tens of K and
tenths of meV, respectively.

\section{Background: Brillouin zone integration for the Green's function} \label{sec:background}

The numerical methods described in this work are based on algorithms
introduced in Ref.~\onlinecite{kaye23_bz} for
the single-particle Green's function
\begin{equation} \label{eq:gfun}
  G(\omega) = \int_\bz d^3\vk \, \Tr \brak{\paren{\hbar\omega - H(\vk) -
  \Sigma(\vk,\omega)}^{-1}}.
\end{equation}
This section summarizes the main conclusions of that work, and may be skipped by readers familiar with it.
We assume $H(\vk)$ is a Hermitian matrix-valued function of $\vk$ which can be evaluated
rapidly, on-the-fly, at arbitrary $\vk$.
This is accomplished using Wannier interpolation based on well-localized Wannier functions in solids~\cite{Marzari/Vanderbilt:1997,Souza/Marzari/Vanderbilt:2001,Marzari_et_al:2012,Marrazzo_et_al:2023}, which has been used previously to efficiently compute challenging BZ integrals~\cite{Wang_et_al:2006,yates07,Lopez_et_al:2012,Wang_et_al:2017,tsirkin21}.
We assume that the dimension of $H(\vk)$ is small enough that the inverse and
trace can also be computed rapidly on-the-fly, for example when considering a downfolded model of physically relevant low-energy states. For cases in which either of these
assumptions do not hold, the resulting bottlenecks tend to take priority over the
resolution of
small scattering rates---our primary concern
here---and other methods may be more appropriate. 

Ref.~\onlinecite{kaye23_bz} analyzes the performance of BZ integration methods for
\eqref{eq:gfun} in terms of a constant scalar-valued scattering rate
$\eta = -\Im \Sigma$.
A similar analysis holds for more general self-energies. 
Uniform integration rules such as the periodic trapezoidal rule (PTR, also
referred to as a Monkhorst--Pack rule \cite{monkhorst76}) converge exponentially, but with a rate proportional to $\eta$,
reflecting the $\OO{\eta}$-scale features in the integrand. Thus 
achieving a fixed accuracy
requires $\OO{\eta^{-1}}$ grid points per dimension, and $\OO{\eta^{-3}}$ points in
total. The simplicity and exponential convergence of the PTR make it
advantageous when $\eta$ is not too small, but it becomes intractable for $\eta$ sufficiently small that the integrand becomes nearly singular
along the surface $H(\vk) = \omega$ (e.g., in Ref.~\onlinecite{kaye23_bz}, thousands of $\vk$ points are required per dimension below the $10$ meV scale).

One strategy to handle this,
explored in Ref.~\onlinecite{henk01} and in the context of optical conductivity in
Ref.~\onlinecite{assmann16}, is to build adaptive grids based on a hierarchical subdivision of the
BZ. In this ``tree-based'' adaptive integration (TAI) approach, the BZ is divided into cells (boxes, or tetrahedra), which are then
automatically refined into smaller cells in regions in which high resolution
is needed according to an error monitor. Ref.~\onlinecite{kaye23_bz} demonstrates that
algorithms of this type have computational complexity $\OO{\eta^{-2}
\log(\eta^{-1})}$ as $\eta \to 0^+$, since they refine logarithmically into cells of diameter
$\OO{\eta}$ along the 2D surface $H(\vk) = \omega$ of area $\OO{1}$.
Another strategy is iterated adaptive integration (IAI), discussed in
Refs.~\onlinecite{henk01,bruno97} and further developed in
Ref.~\onlinecite{kaye23_bz}. To illustrate this in 2D with
$\bz = [-\pi,\pi]^2$, one writes
\begin{gather*}
  \int_{-\pi}^\pi dk_x \, \int_{-\pi}^\pi dk_y \, f(k_x,k_y) =
\int_{-\pi}^{\pi} dk_x \, I_2(k_x), \\
  \mbox{ where} \quad I_2(k_x) \equiv \int_{-\pi}^{\pi} dk_y \, f(k_x,k_y).
\end{gather*}
The outer integral in $k_x$ can be computed by high-order 1D adaptive
integration (see Ref.~\onlinecite[Sec.~3]{kaye23_bz} for a brief description; also
\cite[Sec.~4.7]{press07}, \cite[Sec.~5.6]{kahaner88},
\cite{piessens12}).
At each resulting $k_x$ quadrature node,
$I_2(k_x)$ is in turn evaluated by 1D adaptive
integration with respect to $k_y$. When $f$ is taken to be the integrand
in \eqref{eq:gfun}, both 
$I_2(k_x)$ and $f(k_x, k_y)$ for fixed $k_x$ typically contain localized features at a scale $\OO{\eta}$, and therefore require
only $\OO{\log(\eta^{-1})}$ grid points to resolve using adaptive integration.
In 2D, this yields $\OO{\log^2(\eta^{-1})}$ quadrature nodes in total, and
$\OO{\log^3(\eta^{-1})}$ for the analogous scheme in 3D.

Ref.~\onlinecite{kaye23_bz} concludes that IAI outperforms TAI in the
small-scattering regime, and also describes an optimized implementation of the PTR
for the large-scattering regime. Both the PTR and IAI can be made
high-order accurate, and convergence to a user-specified error tolerance can be
automated, so together these methods allow for the efficient and black-box calculation of the
Green's function in a wide range of settings.

\section{Integration algorithms for the optical conductivity} \label{sec:algorithms}

In this section, we extend these methods to the optical
conductivity \eqref{eq:oc}, emphasizing
points which differ from the case of the
Green's function \eqref{eq:gfun}.
As with the Green's function, we compute $H(\vk)$, its gradient,
and $\mathcal{A}_{\alpha}(\vk)$ on-the-fly using Wannier interpolation~\cite{yates07}. From these, we 
compute the velocity operators $v_{\alpha}(\vk)$ using \eqref{eq:dipole}, and the
spectral function $A(\vk,\omega)$ by direct inversion.
In the case of the Green's function our analysis of the PTR and IAI is based on
the nearly singular behavior of the integrand along the surface $H(\vk) =
\omega$. For the optical conductivity, we must consider the more intricate
structure of the transport distribution $\Gamma_{\alpha \alpha'}(\vk,\omega,\Omega)$. 

\begin{figure}
  \centering
  \begin{tikzpicture}[scale=0.98, every node/.style={transform shape}]
  \node[anchor=south west,inner sep=0](img1) at (0,0) {\includegraphics[width=\linewidth]{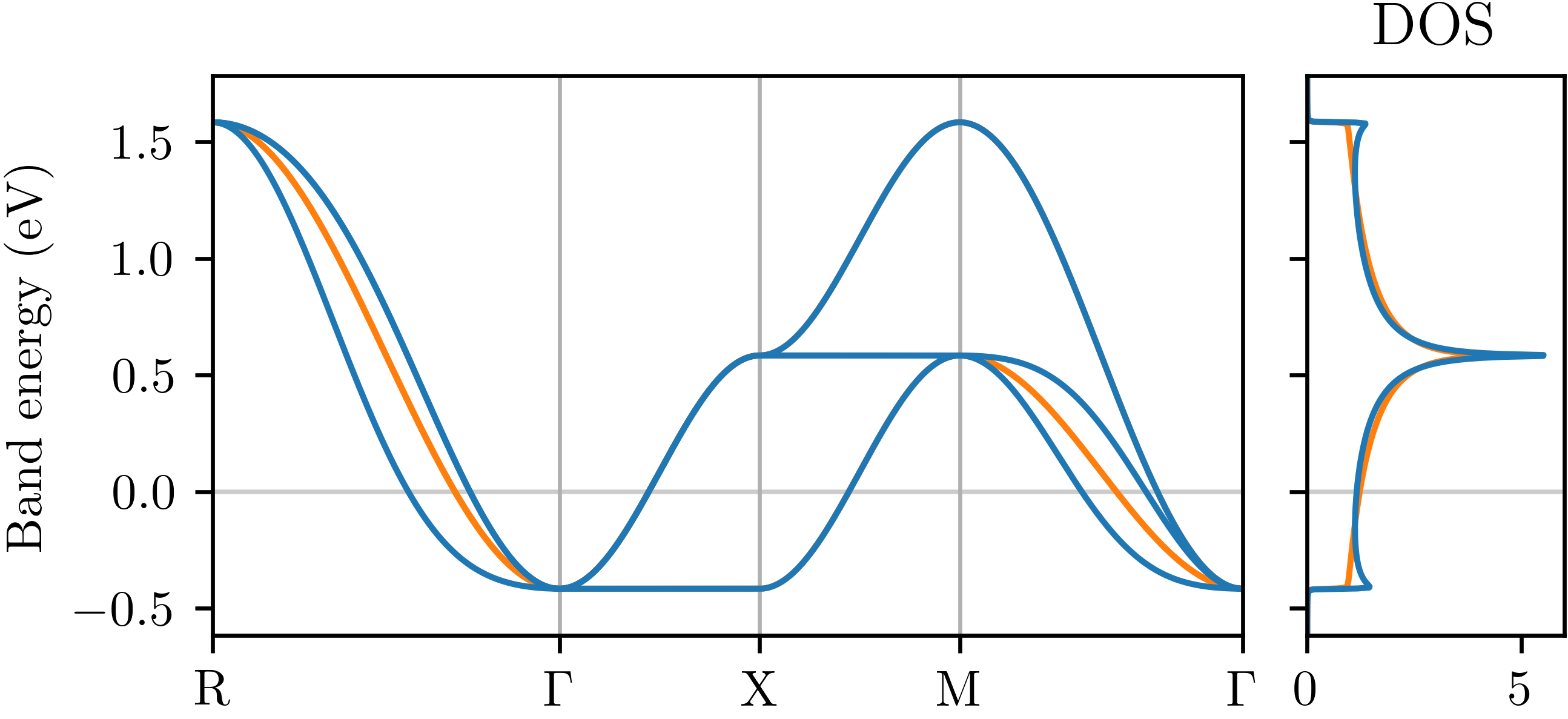}};
    \begin{scope}[x={(img1.south east)},y={(img1.north west)}]
      \node[anchor=south,inner sep=0](img2) at (0.5,1.) {\includegraphics[width=.5\linewidth]{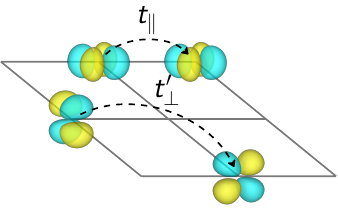}};
      \node[anchor=north] at (.02,1.) {(b)};
      \node[anchor=north] at (.02,1.55) {(a)};
    \end{scope}
  \end{tikzpicture}
  \caption{(a) Schematic illustration of the the tight-binding model (2D slice of
  3D system shown) including hopping terms and exemplary $t_{2g}$-like orbitals. $t_{\parallel}$ and $t'_{\perp}$ correspond to intra-
  and interband hoppings, respectively.
   (b) Band structure and normalized density of states corresponding to the hopping parameters defined in the text (blue lines).
    The orange lines represent the band structure when $t'_{\perp}=0$. The DOS calculation is performed with a broadening of $\eta=1$ meV.
   }
  \label{fig:model}
\end{figure}

In this work we use the model system summarized in Fig.~\ref{fig:model} as an example, but
note that the discussion in this section applies more generally (full details, including numerical calculations, 
will be presented in Sec.~\ref{sec:results}).
It consists of a three-band
tight-binding model on a cubic lattice, motivated by the three-fold degenerate
$t_{2g}$ orbitals, as are present in an octahedral crystal-field splitting.
We define nearest-neighbor intraorbital hopping amplitudes $t_{\parallel}$ for each orbital, as well as an orbital off-diagonal hopping at next-nearest-neighbor distance $t'_{\perp}$ to introduce interband transitions (since these are not allowed by symmetry at nearest-neighbor distance for cubic structures).
For simplicity we do not include an intraorbital next-nearest-neighbor
hopping $t'_{\parallel}$ , although it is typically larger than $t'_{\perp}$ in
realistic systems.
The interorbital hopping results in a splitting of the otherwise degenerate bands in certain regions of the BZ, e.g., on the paths of high symmetry points $R - \Gamma$ and $M - \Gamma$ (orange versus blue lines in Fig.~\ref{fig:model}).
Although the splitting is visible in these low-dimensional paths, predicting the resulting contribution to the interband transitions requires a three-dimensional view of the BZ at multiple transition frequencies.
As will be illustrated in Sec.~\ref{sec:results}, once the optical
conductivity is computed, peaks in the conductivity can be identified, guiding
the visualization of the corresponding contributions in the BZ.

\subsection{Structure of the transport distribution} \label{sec:structure}

We first consider the structure of the integrand $F_\Omega(\omega) \Gamma(\vk,
\omega, \Omega)$ of \eqref{eq:oc} in the $(d+1)$-dimensional $(\vk,\omega)$
space, for fixed $\Omega$.
To facilitate visualization, we allow $d \in \{1, 2, 3\}$. For this discussion, we restrict to
the case of a scalar-valued self-energy $\Sigma = -i \eta$, but our primary
conclusions hold more generally. 
For $\Omega > 0$, we note that $F_\Omega(\omega)$
is a smooth window function which is nearly constant on an interval of width $\OO{\Omega}$, and then decays
exponentially at an $\OO{T}$ rate.
For $\Omega = 0$, it is a localized bump, also decaying exponentially at an $\OO{T}$ rate. In both cases, its role is to truncate the
frequency integral in \eqref{eq:oc} to the \emph{Fermi window}.

With this in mind, we
focus on the structure of the transport distribution
$\Gamma_{\alpha \alpha'}(\vk,\omega,\Omega)$ for fixed $\Omega$ and $\omega$ within the
Fermi window. We will see that it is characterized by peaks of width $\OO{\eta}$ localized
along manifolds
typically of dimensions $d-1$ and $d$ lying in the $(d+1)$-dimensional $(\vk,\omega)$ space. For example, for
$d = 2$, we obtain peaks localized along surfaces (2D manifolds in 3D
$(\vk,\omega)$ space), arising from intraband transitions, and curves (1D
manifolds in $(\vk, \omega)$ space), arising from interband transitions.

\begin{figure*}
  \centering
  \includegraphics[width=1.0\linewidth]{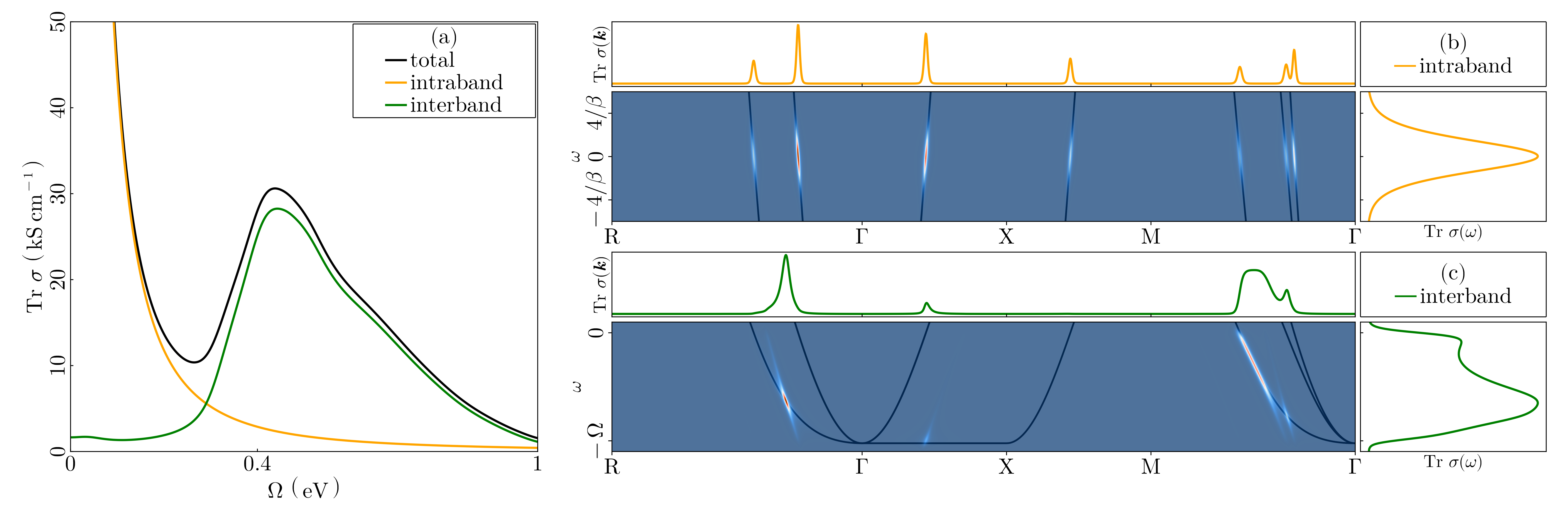}
    \caption{
    Structure of the optical conductivity (a) and its integrand (b,c) for the tight-binding model at temperature
    $T=100$~K and $\eta = 18$~meV. Intraband (b) and interband (c)
    contributions are shown separately (see \eqref{eq:peaks} and the surrounding
    discussion). In (b) and (c), the center subplot shows the spectral
    density of $\Tr\sigma(\vk, \omega)$,
    in color overlayed on the spectrum (black curve), and the
    top and right subplots show the density integrated with respect to $\omega$ and
    $\vk$, respectively.
    Panel (b) shows intraband transitions at $\Omega=0$~eV, localized at the
    Fermi energy on the bands.
    Panel (c) shows interband transitions at $\Omega=0.4$~eV.}
  \label{fig:kpathdensity}
\end{figure*}

For a scalar-valued self-energy, the spectral functions are
diagonal in the eigenbasis of $H(\vk)$, which we refer to as the band basis, and
the transport distribution simplifies:
\begin{equation} \label{eq:peaks}
  \begin{multlined}
    \Gamma_{\alpha \alpha'}(\vk, \omega, \Omega) = \sum_{mn} v_{\alpha,mn}^H(\vk) v_{\alpha',nm}^H(\vk) \\
    \times \frac{\eta^2}{((\omega - \epsilon_{m\vk})^2+\eta^2)((\omega + \Omega-\epsilon_{n\vk})^2+\eta^2)}.
  \end{multlined}
\end{equation}
Here $\epsilon_{n\vk}$ is the $n$th eigenvalue of $H(\vk)$, and
$v_{\alpha,mn}^H(\vk)$ are the elements of the velocity matrix in the band basis,
which vary
slowly, and in particular independently of the $\OO{\eta}$ and $\OO{T}$ length scales we will consider.
For terms with $m \neq n$, the peaks of this quantity in $(\vk,\omega)$ space
correspond to optical transitions of frequency $\Omega$, referred to as interband
transitions. When $m = n$, it is still non-zero for $\eta > 0$, as
allowed by momentum-relaxing electron-electron scattering
processes, and represents an intraband transition \cite{maslov16}.
The total optical conductivity, as well as the intra- and interband contributions for the
tight-binding model, are illustrated in Fig.~\ref{fig:kpathdensity}.

Let us first consider the intraband transitions, $m = n$. When $\Omega$ is
large compared to $\eta$, \eqref{eq:peaks} has two peaks of width $\OO{\eta}$
along the $d$-dimensional manifolds $\omega = \epsilon_{n\vk}$ and $\omega =
\epsilon_{n\vk} - \Omega$ in $(\vk, \omega)$ space. When $\Omega$ is small
compared to $\eta$, these peaks merge into a single larger peak of width
$\OO{\eta}$ along $\omega = \epsilon_{n\vk}$.
In both cases, the contribution
of the intraband transitions arises from peaks in $(\vk, \omega)$ space of width
$\OO{\eta}$ along $d$-dimensional manifolds parallel to the bands. This is
reflective of the $\Omega = 0$, $\eta \to 0^+$ picture, in which intraband
transitions take the form of integrals over Fermi surfaces (i.e., type II in Ref.
\cite{yates07}).
Since the height of each peak 
is $\OO{1/(\Omega^2 + \eta^2)}$, we see that the contribution of the
intraband transitions to the optical conductivity scales as $\OO{\eta/(\Omega^2+\eta^2)}$.

Fig.
\ref{fig:kpathdensity}(b) shows an example of this structure when $\Omega
= 0$ eV. For the tight-binding model described above, we plot the integrand
$\Tr \sigma(\vk, \omega) \equiv F_\Omega(\omega) \Tr \Gamma(\vk,
\omega, \Omega)$ along the path of high
symmetry points (here, the trace is taken over polarization directions). We see that the intraband transitions appear as
peaks of width $\OO{\eta}$ localized along the bands within the Fermi window.
Additionally, the marginal plots show similarly localized features in $\Tr
\sigma(\vk)$ and $\Tr \sigma(\omega)$, which have been integrated with
respect to $\omega$ and $\vk$, respectively.

For interband transitions $m \neq n$, 
we find peaks of width $\OO{\eta}$ along the
manifold on which bands $m$ and $n$ are separated by a distance $\Omega$,
$\abs{\epsilon_{m\vk} - \epsilon_{n\vk}} = \Omega$. Typically, this is a
$(d-1)$-dimensional manifold, although its dimension can be lower (e.g., for optical
transitions occuring at an isolated $\vk$ point, the dimension collapses to $0$)
or equal to $d$ (for parallel bands).
This structure
reflects the $\Omega > 0$, $\eta \to 0^+$ limit, in which interband
transitions take the form of integrals over energy-difference isosurfaces (i.e.,
type III in Ref. \cite{yates07}).
Since these interband features typically occupy one $\eta$-confined dimension less than
intraband features, we expect
their magnitude to be $\OO{1}$ with respect to $\eta$.
An example is shown in
Fig. \ref{fig:kpathdensity}(c) for the same system as above: on the $R - \Gamma$ and $M - \Gamma$ paths, we see peaks of width
$\OO{\eta}$ corresponding to interband transitions within the Fermi window.
Due to finite temperature effects, nearly-parallel bands along the $M - \Gamma$ path contribute a $d$-dimensional feature in $\Tr\sigma(\vk,\omega)$.

For general self-energies, $H(\vk)$ and $\Sigma(\omega)$ are typically not
simultaneously diagonalizable, so intra- and interband transitions cannot be
distinguished as in \eqref{eq:peaks}, and an in-depth analysis is more difficult. 
Nevertheless, our adaptive algorithm remains viable in the general case because the basic
picture of localized peaks in $(\vk,\omega)$ space
remains valid.

\subsection{Periodic trapezoidal rule}

As for the single-particle Green's function, when $\eta$ (or, more generally, $- \Im \Sigma$) is large the
BZ integrals in \eqref{eq:oc} can be efficiently computed using the PTR. As
described in Ref.~\onlinecite{kaye23_bz}, convergence to a specified error
tolerance can be achieved in an automated manner using a global refinement
procedure: the number $N$ of grid points per dimension is
increased by a constant $\Delta N = \OO{1/\eta}$
until a user-defined convergence
threshold is reached. We carry out the integration in a nested manner, taking
the $\omega$ integral as the outer integral, as written in \eqref{eq:oc}, and then, for each fixed quadrature node in
$\omega$, integrating
the transport distribution over $\vk$ using the PTR. For the outer frequency
integral, we use a standard adaptive Gauss--Kronrod quadrature method \cite{quadgk}
(for an explanation of a simple adaptive Gauss quadrature scheme, see
Algorithm 2 of Ref.~\onlinecite{kaye23_bz}).

To understand the computational complexity of this method, we consider the
integrand of the frequency integral in \eqref{eq:oc},
$F_\Omega(\omega) \int_\text{BZ} d\vk \, \Gamma_{\alpha \alpha'}(\vk,\omega,\Omega)$,
the product of the Fermi window function and the $\vk$-integrated transport
distribution. The $\vk$-integrated transport distribution is
a smooth
function of $\omega$ with localized features of width $\OO{\eta}$ arising from
the peaks of the transport distribution itself. The product of these functions has features of
width $\OO{\min(\eta,T)}$ and numerical support in the Fermi window of width
$\OO{\Omega}$.
As a consequence, the adaptive Gauss--Kronrod quadrature can resolve the integral using
$\OO{\log(\Omega/\min(\eta,T))}$ quadrature nodes.
We show in Appendix \ref{app:window} how to evaluate the Fermi window function in
a numerically stable manner, and how to determine its numerical support. 

At each of these frequency quadrature nodes, we perform the $\vk$ integral using the PTR.
Since the localized features of the transport distribution are of width
$\OO{\eta}$, the PTR requires $\OO{\eta^{-d}}$ uniform grid points,
and converges with spectral accuracy thereafter. The total computational
complexity of this method is therefore $\OO{\eta^{-d}
\log(\Omega/\min(\eta,T))}$.
We note that many codes use a
uniform quadrature rule for the frequency integral, leading to an
$\OO{\eta^{-d} \, \Omega/\min(\eta,T)}$ scaling. Simply replacing this with
adaptive frequency integration leads to a significant improvement in efficiency.

The adaptive quadrature method for the frequency integral refines automatically to yield a specified
absolute error tolerance $\varepsilon$. For each $\omega$ point, we also automatically refine the uniform PTR
grid in $\vk$
to achieve $\varepsilon$ accuracy. Given a new
$\omega$ point, we begin with the finest PTR grid obtained from previous $\omega$
points, and refine it as needed. In this way,
we can store the quantities $H(\vk)$ and $v_\alpha(\vk)$ on the finest PTR grid
used thus far, amortizing the cost of computing them over all frequencies. This
refinement strategy is the primary motivation for taking the frequency integral
as the outer integral. Although using this approach could lead to an
over-refined PTR grid for a given
$\omega$ point, in practice we find that the benefit of re-use outweighs this disadvantage. Other more sophisticated
strategies, such as storing these quantities on a sequence of PTR grids, could
be explored as well.

We lastly mention that when the PTR grid is chosen with special lattice
vectors, shifts, and numbers of points along each dimension, the total
number of $\vk$ points can be reduced a factor nearly equal to the number of
point group symmetries~\cite{froyen1989,moreno1992,morgan2020}.
We refer to Ref.~\onlinecite{kaye23_bz} for significantly more detail on the PTR
approach to BZ integration: Section 2 and Appendix D discuss automatic
refinement, Appendix A discusses convergence, Appendix B describes a
irreducible BZ symmetrization procedure, and Appendix C addresses the efficient evaluation of
Wannier-interpolated quantities.

\subsection{Iterated adaptive integration} \label{sec:iai}

When the scattering rate is small, we
favor the IAI scheme discussed in Section \ref{sec:background}. In this case, we
use the adaptive Gauss--Kronrod method for all integrals. We also take the
frequency integral as the inner integral in order to amortize the cost of evaluating
$H(\vk)$ and $v_\alpha(\vk)$ over the number of $\omega$ quadrature points, since unlike for the PTR, the
choice of $\vk$ points is determined on-the-fly. 

The integrand of the inner frequency integral is given for fixed $\vk$ by
$F_\Omega(\omega) \Gamma_{\alpha\alpha'}(\vk, \omega, \Omega)$.
As before, this function contains localized features of width
$\OO{\min(\eta,T)}$ on a window of size $\OO{\max(\Omega,T)}$, so the
cost of the adaptive integration in frequency at fixed $\vk$ is
$\OO{\log(\max(\Omega, T)/\min(\eta,T))}$.

\begin{figure}
  \centering
  \includegraphics[width=\linewidth]{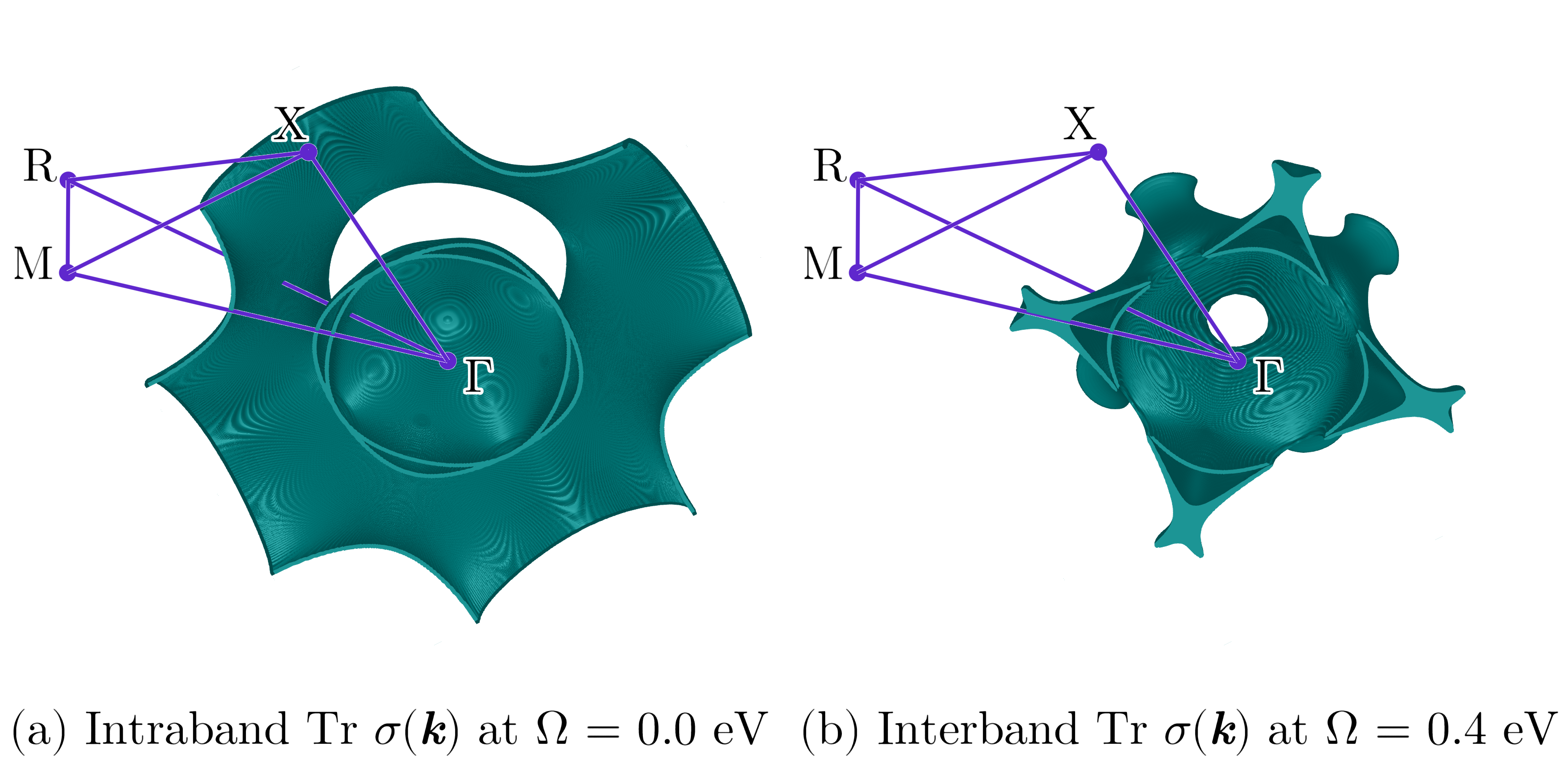}%
   \caption{
   Contributions to the intraband and interband conductivity
   for $T=16$~K at $\Omega= 0$ eV and $\Omega = 0.4$ eV in panels (a) and (b), respectively. We show a
   cross section through the $k_z = 0$ plane of the step function which is colored in the set where $\Tr \sigma(\vk)$ is within 99.5\% of its maximum and vanishes elsewhere, restricted to the first BZ and sampled at 400 $\vk$ points per dimension. At $\Omega=0$ eV the
   integrand
   is highly localized about the Fermi surface, which contains an electron-like
   and a hole-like sheet. The interband integrand is peaked about
   energy-difference isosurfaces in the Fermi window.
   }
  \label{fig:oc_bz}
\end{figure}

The integrand of the outer $\vk$ integral is
$\int_{-\infty}^\infty d\omega \, F_\Omega(\omega) \Gamma_{\alpha \alpha'}(\vk, \omega,
\Omega)$.
In general, this quantity inherits the $\OO{\eta}$-scale features of the transport distribution.
As described in Sec.~\ref{sec:structure}, intraband transitions at $\Omega=0$ give rise to
$d$-dimensional $\OO{\eta}$-scale peaks along the bands in the Fermi window.
After integrating in $\omega$, we obtain localized features in the $\vk$
integrand where a band crosses the Fermi window, as shown in
Fig.~\ref{fig:oc_bz}(a), which resembles the Fermi surface.
These features are of size $\OO{\max(\eta,T)}$, since this is the scale over
which the Fermi window truncates the intraband contributions.
This leads to an 
$\OO{\log^d(\max(\eta,T)^{-1})}$ scaling in the number of $\vk$ points required
by IAI, assuming the error tolerance is scaled according to the $\OO{1/\eta}$
magnitude of the Drude peak.

On the other hand, interband transitions at finite $\Omega$ in
$(d+1)$-dimensional $(\vk,\omega)$ space typically give rise to $(d-1)$-dimensional
$\OO{\eta}$-scale features.
The integrand in $\vk$ inherits localized features at scales $\OO{\eta}$ and
$\OO{T}$ appearing in the inner frequency integral, as shown in
Fig.~\ref{fig:oc_bz}(b). The features between $M - \Gamma$ and $R - \Gamma$
correspond to those in Fig.~\ref{fig:kpathdensity}(c), although the structure of
those peaks away from the high-symmetry lines is more intricate.
In order to resolve both length scales, IAI will adaptively refine to the smallest,
so its cost will scale as $\OO{\log^d(\min(\eta, T)^{-1})}$.

Combining these estimates, we expect an $\OO{\log^d(\max(\eta,T)^{-1})
\log\paren{\frac{\max(\Omega, T)}{\min(\eta,T)}}}$ computational complexity if we restrict to intraband
transitions, and $\OO{\log^d(\min(\eta,T)^{-1}) \log\paren{\frac{\max(\Omega,
T)}{\min(\eta,T)}}}$ in general. For example, for the 
Fermi liquid scaling $\eta \propto T^2$, we expect
$\OO{\log^{d+1}(\eta^{-1})}$ cost scaling at low
temperatures. Thus, in the typical case, 
we find a complexity reduction of $\OO{(\eta^{-1}/\log \eta^{-1})^d}$ compared
with the PTR, similar to that for the single-particle Green's function.

We note that the magnitude of intermediate integrals might scale differently
with $\eta$ than the magnitude of the total integral, possibly increasing the
overall computational complexity if absolute error tolerances are used. In
principle, for efficiency, intermediate error tolerances should be scaled
accordingly to attain a fixed number of significant digits at each step.
The robust use of such relative error tolerances is a topic of our
ongoing research. Here, we prioritize robustness and use a fixed absolute error
tolerance, except for the inner frequency integral, for which
the tolerance is scaled as $\OO{1/\eta}$ corresponding to its expected magnitude
at optical transitions.

When $\eta$ is very small, the integrand
typically contains tall, narrow peaks which might fall between the initial coarse
quadrature grid points of the adaptive integration algorithm, leading it to
falsely conclude that convergence has been reached. In Appendix \ref{sec:peak},
we analyze this ``peak-missing'' problem in detail. We propose a simple
solution, called the auxiliary integrand method, which addresses the problem by
including a simpler integrand---for instance, the Green's function---in the
integration procedure to guide the selection of adaptive grid points at little
additional cost.

Finally, we mention an optimization for scalar-valued self-energies in both
the PTR and IAI: $H(\vk)$ can be diagonalized once per $\vk$-point to compute the transport
distribution in the band basis, as in \eqref{eq:peaks}, thereby replacing on-the-fly matrix inversions with matrix multiplications.

\section{Numerical results}
\label{sec:results}

\subsection{Tight-binding model}
\label{subsec:tb_model}
The Hamiltonian for the tight-binding model introduced
at the beginning of Sec.~\ref{sec:algorithms} is given by
\begin{align} \label{eq:model}
    \hat H = \sum_{<ij>}\sum_{\alpha} t_{i-j}^{\alpha} \hat{c}_{i,\alpha}^\dagger \hat{c}_{j,\alpha} &+ \sum_{\ll ij\gg}\sum_{\alpha\neq\alpha'} t_{i-j}^{\alpha\alpha'} \hat{c}_{i,\alpha}^\dagger \hat{c}_{j,\alpha'},
    \\ t_{\pm e_{\alpha'}}^{\alpha} &= t_\parallel (1-\delta_{\alpha\alpha'}), \nonumber
    \\ t_{\pm(e_\alpha + e_{\alpha'})}^{\alpha\alpha'} &= -t_{\pm(e_\alpha - e_{\alpha'})}^{\alpha\alpha'} = t'_\perp. \nonumber
\end{align}
Here, $i, j$ are site labels on the cubic lattice, $\alpha,\alpha'\in\{x,y,z\}$,
$e_\alpha$ is a lattice vector in direction $\alpha$, and $\hat{c}_{i,\alpha}$
is a spinless fermion annihilation operator in an orbital polarized along direction
$\alpha$ at site $i$.
We use single (double) angle brackets to denote summation over nearest
(next-nearest) neighbor sites, and only the non-zero hopping amplitudes are indicated.
We choose $t_{\parallel} = -0.25$ eV, resulting in a bandwidth of 2 eV, and $t'_{\perp} = 0.05$ eV.
The density of states (DOS), shown in Fig.~\ref{fig:model}~(b), is computed using IAI as in
Ref.~\onlinecite{kaye23_bz}.
We use a Fermi liquid scaling for the temperature dependence of the scattering rate, i.e. $\eta = \frac{k_B \pi}{Z T_0} T^2$, with the quasiparticle renormalization chosen to be $Z=0.5$ and the characteristic Fermi liquid temperature scale $T_0 = 300$ K, corresponding to typical values of a moderately correlated metal.
We use the self-energy $\Sigma = -i\eta$.
We set the filling to be one electron distributed across the three bands, i.e.,
we choose $\mu(T)$ so that $\hbar \int d\omega f(\omega-\mu/\hbar) \operatorname{DOS}(\omega) = 1$.
We note that the second
term in \eqref{eq:dipole} does not contribute to the optical conductivity because
$\mathcal{A}_\alpha$ is proportional to the identity in the chosen model.
All BZ integrals are computed on the irreducible Brillouin zone (IBZ), as described in Appendix \ref{app:ibzlimits}.
We perform all calculations using the Julia library
\texttt{AutoBZ.jl}~\cite{Van-Munoz/Beck/Kaye:2024}, which implements the algorithms described here and in Ref.~\onlinecite{kaye23_bz}.

\subsection{Timing results}

\begin{figure}
  \centering
  \includegraphics[width=\linewidth]{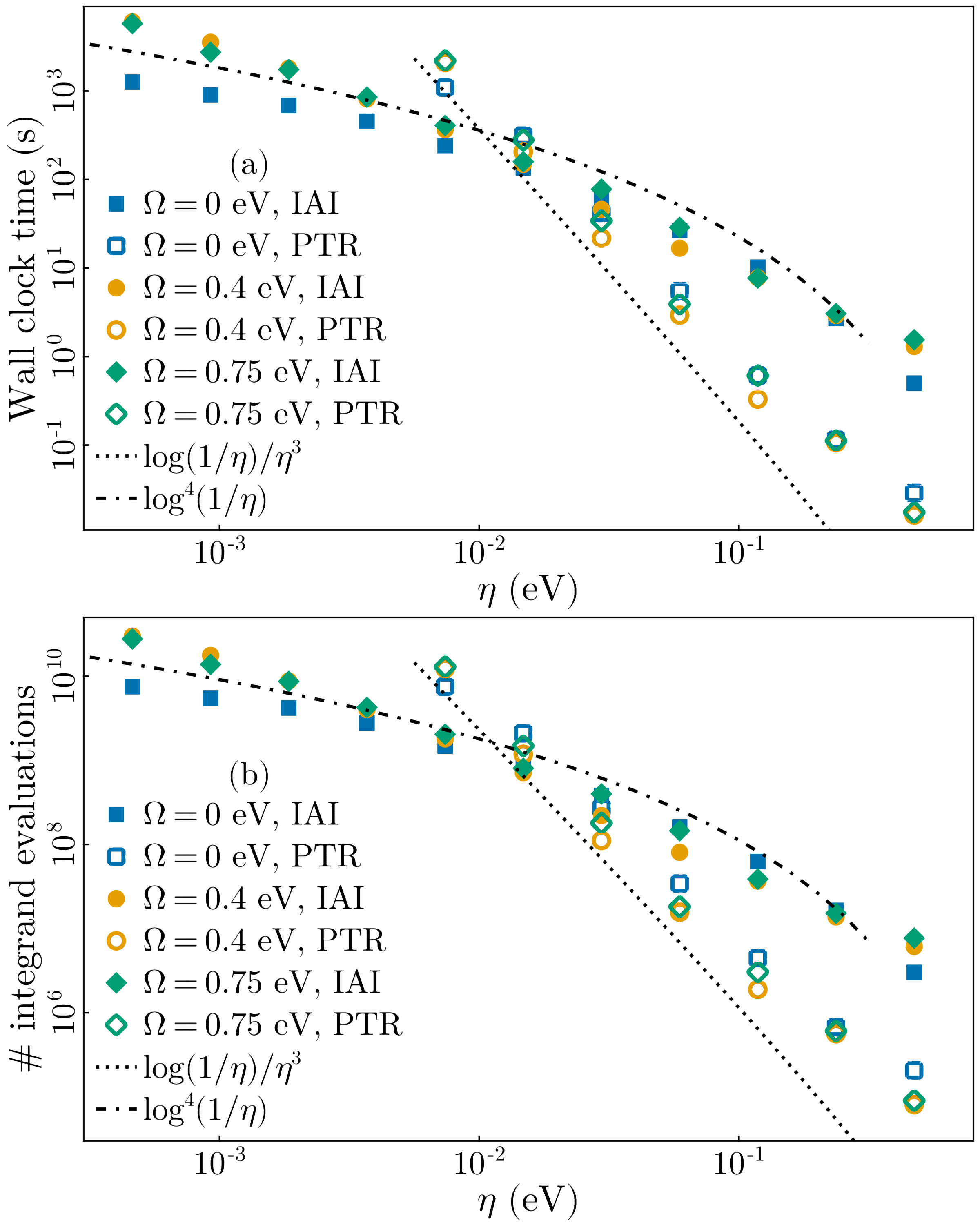}%
   \caption{Wall clock time (a) and number of integrand evaluations (b) versus $\eta$,
     for the calculation of the optical conductivity of the example tight-binding model Hamiltonian.
   }
  \label{fig:crossover}
\end{figure}

Fig.~\ref{fig:crossover} presents the wall clock timings and number of integrand
evaluations required by the PTR and IAI methods
to compute the DC conductivity $\Omega=0$ eV (blue squares) on a single core of a workstation
with an Intel i5-12500 processor and 64 GB of RAM. We choose integration tolerances to give
five significant digits of accuracy, with absolute tolerances
chosen relative to a fit to the asymptotic scaling obtained from several
large-$\eta$, PTR-based estimates of the magnitude of the integral.
For IAI (filled symbols), we observe the expected $\OO{\log^4(\eta^{-1})}$ cost scaling, consistent with the
Fermi liquid scattering rate $\eta \propto T^2$.
The PTR (open symbols) also displays the expected $\OO{\eta^{-3}\log(\eta^{-1})}$ cost scaling.

For $\Omega > 0$ (orange circles and green diamonds), we observe the same cost scaling for the PTR as in the
$\Omega=0$ case. For IAI, we observe a slightly larger scaling than expected (though still sub-algebraic),
and suspect the discrepancy is related to
the choice of intermediate error tolerances which are
kept constant for the $\vk$ integrals.

We observe that for $\eta \lessapprox 0.01$ eV, IAI is more efficient than the PTR in
several single-frequency calculations.
For the most challenging calculation, with $T = 16$ K, $\eta = 0.46$ meV, and
$\Omega = 0.4$ eV, IAI requires 1.68 core-hours, whereas we estimate that the PTR
would require 1750 core-hours for similar accuracy.
Table~\ref{tab:nptr} shows the number of grid points used per dimension by the PTR for
various values of $T$ and $\eta$.
We were unable to complete the PTR calculation for
$T \leq 45$ K due to inadequate memory; at $T=45$~K approximately 2500 $\vk$
points per dimension would have been required.

\begin{table}
  \centering
  \setlength{\tabcolsep}{4.0pt}
  \renewcommand{\arraystretch}{1.4}
  \begin{tabular}{|c|c|c|c|c|c|c|c|c|c|} 
   \hline
    $T$ (K)     & 1024 & 512 & 256 & 128 & 64 \\ \hline
    $\eta$ (eV) & 1.9 & 0.47 & 0.12 & 0.030 & 0.0074 \\ \hline
    $N$         & 10 & 27 & 83 & 335 & 1048 \\
   \hline
  \end{tabular}
  \caption{Maximum number $N$ of PTR grid points per dimension used for any
  frequency $\Omega$ at each temperature $T$ presented in Fig.
  \ref{fig:crossover}. The error tolerance for the automatic
  algorithm is chosen to give five significant digits.}
  \label{tab:nptr}
\end{table}

\subsection{Interband transitions and \texorpdfstring{$T^2$}{quadratic} resistivity in a Fermi liquid}

Accurately resolving scaling laws and the substructure of (interband) peaks in an ab-initio setting is computationally challenging, with existing codes typically reliant on the PTR and limited to $\eta$ at minimum tens of meV for resolved results, i.e. $T$ roughly $100$~K with the parameters chosen here.
Our IAI method allows us to resolve materials properties such as the DC
conductivity or interband features at temperatures as low as $T=16$~K, a largely
unexplored regime in finite-temperature electronic structure calculations.

\begin{figure}
  \centering
  \includegraphics[width=1.0\linewidth]{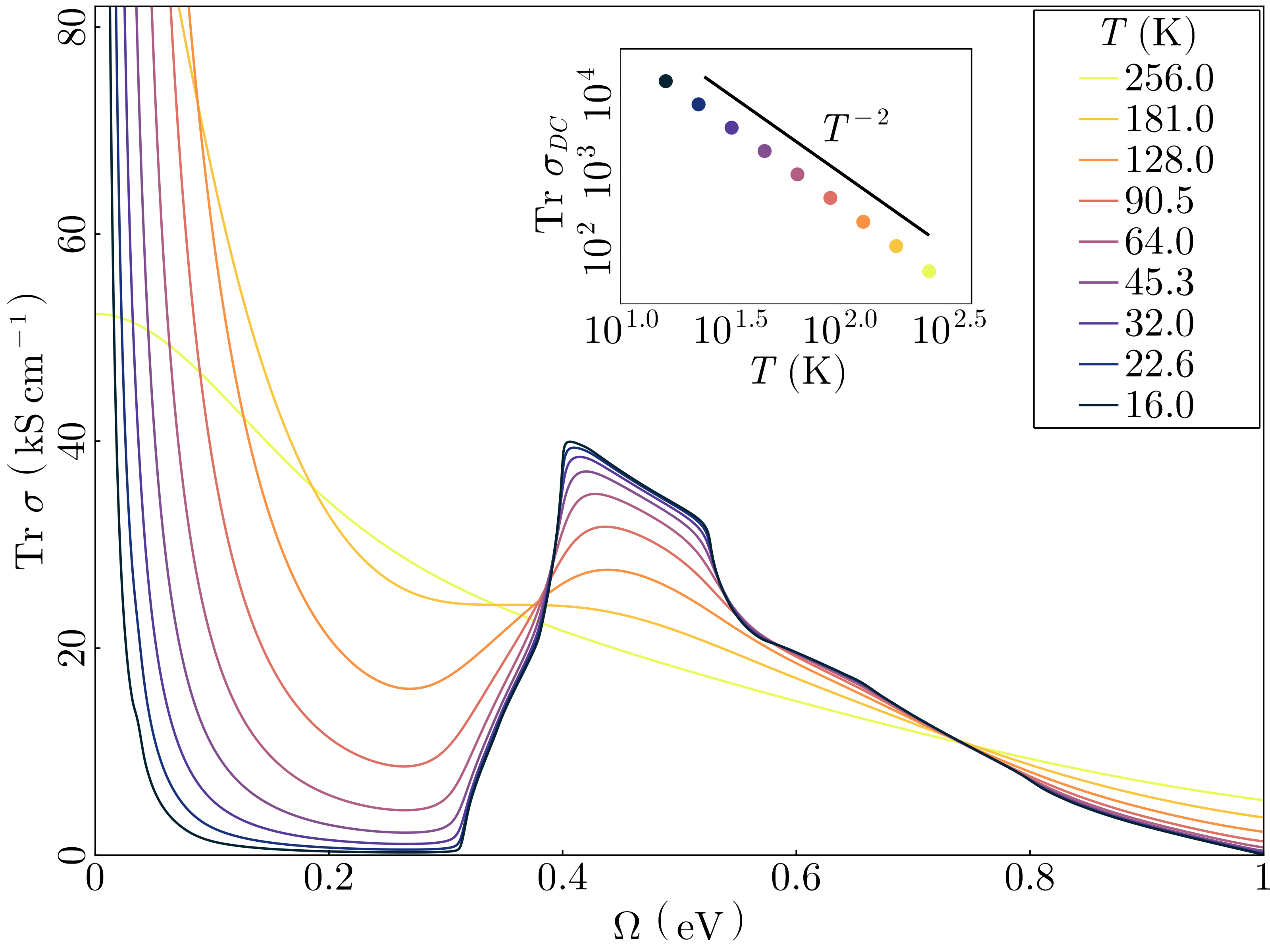}
    \caption{
    Optical conductivity for the model Hamiltonian described in Sec.~\ref{subsec:tb_model}, with the
    inset demonstrating the expected $1/T^2$ scaling of the DC conductivity.
    }
  \label{fig:oc_fermiliquid}
\end{figure}

In Fig.~\ref{fig:oc_fermiliquid} we plot the real part of the optical conductivity for the tight-binding model introduced in Sec.~\ref{subsec:tb_model} with a Fermi liquid scaling of the scattering rate, and
temperatures ranging from 256 K to 16 K.
This corresponds to scattering rates from 0.12 eV to 0.46 meV.
We computed the conductivity using adaptive piecewise Chebyshev interpolation in $\Omega$, as in Sec. 5 of Ref.~\onlinecite{kaye23_bz}, with an error tolerance of three significant digits, and an error tolerance of five significant digits for IAI at each queried $\Omega$. For $T=16$~K, $633$ values of $\Omega$ were queried by the algorithm in total.

We emphasize three key results.
First, the DC conductivity $\sigma(0)$, an observable typically strongly
affected by underresolved BZ sampling, scales as $1/T^2$, following the expected
contribution from quasiparticle-quasiparticle scattering in a Fermi liquid \cite{yamada86}.
Second, the Drude peak at finite frequency associated with intraband transitions narrows with a FWHM scaling of $T^2$ for decreasing temperature (not shown here), a consequence of the optical sum rule conserving the particle number.
Third, the broad absorption peak due to the interband hoppings at $\approx 0.4$
eV becomes refined as the temperature is decreased, revealing details of the
underlying electronic structure beyond approximate peak positions in the $T \to
0^+$ limit.

\section{Conclusion} \label{sec:conclusion}

We have presented two general purpose and high-order accurate integration
algorithms for the calculation of the optical conductivity, intended for
settings in which the transport distribution, and in particular the Green's function, can be evaluated rapidly on-the-fly using Wannier interpolation. In
both cases, we use adaptive frequency integration. For BZ integration, we
recommend the PTR approach for larger scattering rates, $\eta \gtrapprox
0.01$~eV, and the IAI approach otherwise.
In IAI, the 4D $(\omega,\vk)$ integral is computed
by nested one-dimensional adaptive integration, leading to
polylogarithmic computational complexity with respect to the inverse of the scattering
rate. With some modifications, our approach can be extended to response functions besides
the optical conductivity.

These algorithms allow, for the first time, reliable and automated access to sub-meV energy scales
in finite-temperature electronic structure calculations in modest compute times.
They offer robustness, error control, and efficiency both in applications of finite temperature methods such as dynamical mean-field theory~\cite{Georges_et_al:1996},
in which a self-energy provides quasiparticle scattering rates, and in ground-state calculations,
in which scattering is typically added artificially.
In particular, our algorithms automate convergence testing which, if neglected, can lead to
severe spectral artifacts.
As such, they can
be used to provide accurate benchmarks for comparison with experimental spectra,
as well as in high-throughput screenings and machine
learning of materials properties.

The methods described here have been implemented in the open source Julia package
\texttt{AutoBZ.jl}~\cite{Van-Munoz/Beck/Kaye:2024},
which can be used for a broad class of BZ integrals.

\acknowledgments

We thank A. Millis, J. Mravlje and C. Dreyer for helpful discussions. We are
especially grateful to S. G. Johnson and F. Kugler for feedback on our implementation.
The Flatiron Institute is a division of the Simons Foundation. 

\appendix

\section{The peak-missing problem and auxiliary integration} \label{sec:peak}

This section describes a technical improvement to the basic IAI algorithm
described in Sec.~\ref{sec:iai} which is necessary to obtain robustness for
very small values of $\eta$. Conceptually, the problem is as follows. Adaptive
Gauss quadrature algorithms work by applying a fixed $p$-node quadrature rule to
the subintervals of integration, estimating the error on those subintervals
(e.g., in the case of an adaptive Gauss--Kronrod scheme, by comparing the result
to an alternative result obtained using an embedded higher-order quadrature
scheme), and subdividing them if the error estimate is larger than a user-determined
threshold. Since a $p$-node Gaussian quadrature rule has order of
accuracy $2p$, a choice like $p = 4$ might suffice for typical requested accuracies. In
this case, a highly localized feature
may fall between 
the quadrature nodes of the initial subintervals,
leading to small error estimates which cause the algorithm to terminate
before these features are resolved. We therefore refer to this failure
mode as the ``peak-missing'' problem.

More precisely, the peak-missing problem arises from a relationship between the
magnitude of the integral and the magnitude of the tails of its integrand.
We illustrate
this for the optical conductivity at $\Omega = 0$, where the problem is most
severe, using a simplified but illustrative model of the transport distribution:
\[\Gamma(\omega) = \frac{\eta^2}{((\omega-c)^2 + \eta^2)^2}.\]
If we replace the Fermi function by a characteristic function on $[-1,1]$, then
the inner frequency integral in IAI becomes $\int_{-1}^1 d\omega \,
\Gamma(\omega)$. A direct calculation shows that this is $\OO{\eta^{-1}}$ when $\abs{c} < 1$,
consistent with the typical magnitude of the DC conductivity.
However, $\Gamma(\omega)$ decays rapidly, as $\OO{\eta^2/\omega^4}$, so it is of
magnitude $\OO{\eta^2}$ a distance $\OO{1}$ from the peak at $x = c$. As a concrete example, let
us take $c = 0.17$, $\eta = 10^{-3}$, and $p = 4$. Then, the closest node to the peak at $x = c$ is at
$\omega \approx 0.34$, and we have
$\Gamma(0.34) \approx 0.0012$. In other words, all quadrature nodes are far from
the peak, which has magnitude $\Gamma(0.17) = 10^6$.
The Gauss quadrature rule estimates the integral as $\approx 7.9 \times 10^{-4}$, and the
error estimate obtained by comparing this rule to the corresponding embedded higher-order Gauss--Kronrod rule is
$\approx 4 \times 10^{-4}$. The correct integral, on the other hand, is $\approx 1571$. For
a user-requested error tolerance of $\varepsilon = 10^{-2}$, giving approximately
five relative digits of accuracy compared with the expected $\OO{\eta^{-1}}$
magnitude of the result, the algorithm will find that $4 \times 10^{-4} <
\varepsilon$ and terminate erroneously at the first step.

To avoid termination of the algorithm before the peak is resolved, we are forced to scale
$\varepsilon$ as $\OO{\eta^2}$. Indeed, the initial error estimate is obtained
from a quadrature rule consisting of nodes which might be a distance $\OO{1}$
from the peak, yielding an estimate of magnitude $\OO{\eta^2}$.
This is an undesirable scenario, since a user-specified accuracy should be delivered
robustly regardless of the properties of the integrand. Even if we
accept this restriction,
choosing $\varepsilon = \OO{\eta^{-2}}$ yields $\OO{\eta^{-3}}$ significant
digits, typically far more than is needed in applications, resulting in unnecessary computational work.

A simple solution would be to force the algorithm to begin with a finer collection of
uniformly-spaced sub-intervals. However, this simply pushes the problem to
slightly smaller values of $\eta$, at the cost of an unnecessary uniform refinement of the
interval---it does not eliminate the $\varepsilon = \OO{\eta^2}$ scaling.

To solve the problem, we introduce \emph{auxiliary integrands}
$I_j(\omega)$ with the following properties:
\begin{itemize}
  \item The functions $I_j$ are smooth, and at least one has a peak in the
  location of each peak of $\Gamma$.
\item When integrated simultaneously,
  the $I_j$ do not have a peak-missing
  problem.
  \item Ideally, simultaneous adaptive integration of the functions $I_j$ does not require a finer quadrature
  grid than adaptive integration of $\Gamma$ itself.
  \item Ideally, evaluating the $I_j$ is no more expensive than evaluating $\Gamma$ itself.
\end{itemize}
For example, in the case of the transport distribution at $\Omega = 0$, $\Gamma(\vk, \omega,
0)$, $I_1(\omega) = F_\Omega(\omega) \Tr G(\vk,\omega)$ satisfies all of
the required properties: its integral is $\OO{1}$ in
magnitude, as are
its tails, and computing it is already a step of computing the transport distribution. 
We can then perform IAI simultaneously on the integrand and the
auxiliary integrands, requiring (in addition to the ordinary procedure) that a
global error tolerance $\varepsilon_{\text{aux}}$ be satisfied for each auxiliary
integrand in order to terminate subdivison of integration subintervals. The
auxiliary tolerance $\varepsilon_{\text{aux}}$ can be chosen large and
independent of the tolerance $\varepsilon$, since the role of the auxiliary integration
procedure is only to identify the location of localized features in the
integrand, not to accurately resolve these features. Assuming the auxiliary integral is
$\OO{1}$, we typically choose $\varepsilon_{\text{aux}} = 0.01$~eV$^{-1} \times
\det(B)$, where $B$ is a matrix containing the reciprocal lattice vectors as columns.

For $\Omega > 0$, choosing an auxiliary integrand is less straightforward, and
we describe two possible approaches. The first is to use $I_1(\omega)$ as for
the $\Omega = 0$ case 
and $I_2(\omega) = F_\Omega(\omega) \Tr G(\vk,\omega+\Omega)$. However, since
the interband transitions are localized only at band intersections, this choice
of auxiliary integrands, while conservative, does not satisfy the third
criterion above and causes refinement in locations at which the transport
distribution itself is not localized.

The second is to use an ``artificial
smearing'' approach; we choose $I_j(\omega) = \Gamma_{\eta_j}(\vk, \omega,
\Omega)$, where $\Gamma_\eta$ is identical to the transport distribution but
with $\Sigma$ replaced by $\Sigma - i\eta$ for $\eta > 0$. We then choose a
sufficiently dense sequence of smearing parameters $\eta_j$ to avoid
peak-missing. The peaks of these auxiliary integrands are at the same locations as
the transport distribution, but they are less localized, thereby avoiding the
peak-missing problem. A logarithmically-spaced sequence of smearing parameters,
such as $\eta = 0.1, 0.01, \ldots$, is typically sufficient, and the total
computational cost is multiplied by the $\OO{\log(\eta^{-1})}$ number of
auxiliary smearing parameters.

For $\Omega > 0$, we use the first approach, applied only to the
inner frequency integral. We find that this is typically sufficient to avoid
peak-missing and less
expensive than the second approach. Although it is reasonably straightforward in
practice to detect and solve a peak-missing problem by switching on an auxiliary integrand,
further exploring and automating optimal strategies is a topic of our ongoing research.

\section{Frequency-dependent self-energies} \label{app:seinterp}

Although we have presented examples of optical conductivity calculations using
frequency-independent self-energies, in general $\Sigma(\omega)$ must be
evaluated at arbitrary points within the adaptive frequency integration
algorithm used by both the PTR and IAI methods. For the IAI method, in which we
take the frequency integral as the inner integral, this is a
performance-critical step. We recommend dividing the frequency interval into a dense
collection of uniformly-spaced subintervals, and representing $\Sigma(\omega)$
by a low-order Chebyshev polynomial interpolant on each subinterval.
$\Sigma(\omega)$ can then be evaluated inexpensively on-the-fly at a given point
$\omega$ by evaluating the interpolant corresponding to the sub-interval
containing $\omega$, either using the Chebyshev
recurrence relation or barycentric interpolation \cite{berrut04}. The number of subintervals should be taken
large enough to represent $\Sigma(\omega)$ to well within the IAI error
tolerance. An adaptive collection of subintervals can also be used for highly
localized self-energies, but this increases the cost of locating the subinterval
containing an evaluation point (e.g. by descending a binary tree), and yields
only a memory savings.

To construct such a representation, we must evaluate $\Sigma(\omega)$ at Chebyshev
nodes on each subinterval. In some cases, $\Sigma$ can be computed at arbitrary
points of interest, for example if it is given as the output of analytic
continuation methods such as
Nevanlinna analytic continuation \cite{fei21,fei21_2,iskakov23} or
rational interpolation-based methods \cite{Vidberg/Serene:1977,huang23,ying22,zhang23}.
In other cases, it may be necessary to interpolate from data at
given nodes to the Chebyshev nodes, e.g., in the maximum entropy method~\cite{Gubernatis_et_al:1991,Wang_et_al:2009}
or the numerical renormalization group approach to quantum impurity problems~\cite{Wilson:1975}.
For such cases, the method of choice is problem-dependent (high-order local polynomial interpolation and AAA rational
interpolation \cite{nakatsukasa18} are possibilities). This conversion step is a precomputation, and should be
carried out
as accurately as possible.

\section{Numerical treatment of the Fermi window function} \label{app:window}

\subsection*{Numerically stable evaluation}

Direct evaluation of the Fermi window function $F_\Omega(\omega)$ as defined in
Sec. \ref{sec:intro} is numerically unstable, due to catastrophic cancellation
in the difference of Fermi-Dirac functions. To simplify the discussion, we
non-dimensionalize by taking $x \gets \beta \omega$, $X \gets \beta
\Omega$, and write $f(x) = (\exp(x)+1)^{-1}$, $F_X(x) = (f(x)-f(x+X))/X$. 
We have
\begin{equation} \label{eq:win}
  F_X(x) = \frac{\tanh(X/2)/X}{1 + \frac{\cosh(x+X/2)}{\cosh(X/2)}}.
\end{equation}
While this formula does not suffer from catastrophic cancellation, evaluating
the hyperbolic cosines might cause numerical overflow even though their ratio is of moderate
magnitude, in particular at low temperatures. To address this, we use the formula
\begin{equation}
  \frac{\cosh(x+X/2)}{\cosh(X/2)} = e^{\abs{x+X/2}-X/2}\frac{1 + e^{-2\abs{x+X/2}}}{1 + e^{-X}}
\end{equation}
for cases in which overflow would occur.  Overflow of the evaluation of the
exponential in this expression only occurs at values in which underflow in \eqref{eq:win} is
acceptable.

\subsection*{Truncation}

In order to truncate the frequency integral in \eqref{eq:oc} to within a
user-specified error tolerance $\varepsilon$, we must determine the numerical support of the
Fermi window function. To do this, we truncate when $\abs{F_X(x-X/2)} \leq
\varepsilon$, with the shift centering the window function. A simple
calculation, using only the inequality $1 + \cosh(x+X/2)/\cosh(X/2) > (e^{\abs{x+X/2}}/2)/\cosh(X/2)$,
leads to
\begin{equation}
  \label{eq:trunc}
  \abs{x+X/2} \geq \log\paren{\frac{2\sinh(X/2)}{\varepsilon X}} = \frac{X}{2} + \log\paren{\frac{1-e^{-X}}{\varepsilon X}}.
\end{equation}
This yields a simple criterion for truncating the frequency integral,
which depends only on $\beta$, $\Omega$, and $\varepsilon$, i.e., given $X =
\beta \Omega$ and $\varepsilon$ then \eqref{eq:trunc} gives a lower bound on $x$ for which
\eqref{eq:win} is bounded above by $\varepsilon$. We note that whereas the
first expression on the right hand side might overflow, the second is
numerically stable if one evaluates $1-e^{-X}$ using the \texttt{expm1} function \cite{expm1}.

\section{Iterated adaptive integration on the irreducible Brillouin zone} \label{app:ibzlimits}

Ref.~\onlinecite{kaye23_bz} provides a detailed algorithm to reduce the PTR to
the IBZ for systems with non-trivial point group symmetries. It also points out
that for IAI, one can simply integrate over the IBZ rather than the full
BZ,
\begin{equation*}
  \begin{multlined}
    \int_{\text{IBZ}} dk_x \, dk_y \, dk_z = \int_{a_1}^{b_1} dk_x \,
\int_{a_2(k_x)}^{b_2(k_x)} dk_y \, \int_{a_3(k_x,k_y)}^{b_3(k_x,k_y)} dk_z,
  \end{multlined}
\end{equation*}
where the limits of integration are determined by the boundary of the IBZ. Then
the optical conductivity can be obtained from the integral
$\sigma_{\alpha \alpha'}^{\text{IBZ}}$ restricted to the IBZ by applying the
symmetries $S$ of the point group $G$, represented as orthogonal matrices:
\begin{align}
  \sigma_{\alpha \alpha'} = \sum_{S \in G} S_{\alpha \gamma} \sigma_{\gamma \gamma'}^\text{IBZ} S_{\gamma' \alpha'}^{-1}.
\end{align}
This follows from the definition of the Bloch Hamiltonian, $\hat{H}_{\vk} =
e^{-i\vk\cdot\hat{\bm{r}}} \hat{H} e^{i\vk\cdot\hat{\bm{r}}}$, the cyclic
property of the trace in \eqref{eq:tdist}, and the observation that a rotation of
$\vk$ is equivalent to a rotation of the position operator, which
ensures $\Gamma(S\vk, \omega, \Omega) = S\,\Gamma(\vk, \omega, \Omega) S^{-1}$
as long as the self-energy is not momentum-dependent.
Here, we describe how to determine the IBZ limits of
integration on-the-fly for any point group symmetry.

The IBZ can be described as a convex polyhedron, and the first step is to obtain
some description of this polyhedron from the crystal lattice vectors and atomic
positions. Ref.~\onlinecite{jorgensen22} describes an algorithm for this
task, returning the IBZ as a convex hull object which can either be described as
an intersection of half-spaces, or a collection of vertices and faces  (see also
Ref.~\onlinecite{stokes05}). This algorithm is implemented in the open-source Julia package
\texttt{SymmetryReduceBZ.jl} \cite{symmetryreducebz}.

We begin from the representation of the convex polyhedron as a collection of
vertices and faces. The faces are described by ordered subsets of these
vertices, from which the edges can be determined as line segments. The outer limits
$(a_1, b_1)$ are given by the maximum and minimum $k_x$ coordinates of
the vertices. Then the integral in $k_x$ can be performed, and the
limits $(a_2(k_x), b_2(k_x))$ of the $k_y$ integral must be determined for each
fixed quadrature node $k_x$. This coordinate determines a polygon given by the
vertices at the intersection of the plane of constant $k_x$ with the edges of
the faces of the polyhedron. The limits $(a_2(k_x), b_2(k_x))$ are the maximum and
minimum $k_y$ coordinates of these vertices. Finally, for fixed $k_x$ and $k_y$,
the limits $(a_3(k_x,k_y), b_3(k_x,k_y))$ can be similarly determined from
the intersection of the line of constant $k_x$ and $k_y$ with the polygon's edges.

We note that the presence of vertices in the polygonal IBZ can cause the inner
integrals in IAI to fail to be continuously differentiable, interfering with
high-order convergence. This is straightforwardly remedied by adding the vertices of the IBZ as break
points in the adaptive quadrature scheme.

We also note that one cannot expect, in general, that integrating over the
IBZ leads to a reduction in computational cost proportional to the number of
group symmetries. Indeed, the cost of adaptive integration is mainly determined
by the number of localized features in the integrand over the integration
domain, which might not be reduced in the IBZ compared to the full BZ. In fact,
in certain pathological cases, for example involving a continuous translation
symmetry along a particular dimension, integrating over the IBZ could introduce
additional features in the middle and outer integrals, due to the dependence on
the limits of integration themselves. Furthermore, the computational cost could
depend on the particular choice of the irreducible wedge of
the BZ and the order of integration, and determining the optimal choice is not straightforward a priori. However, in many typical cases, integrating
over any choice of the IBZ leads to a substantial reduction in computational cost.

\bibliography{autobz_oc}
\end{document}